\documentclass[aps,12pt,preprintnumbers]{revtex4}

\usepackage{graphicx}
\usepackage{amsmath,amssymb,epsfig,bm,comment}
\usepackage{fancyhdr}
\usepackage[dvips]{color}
\usepackage{mathrsfs}
\usepackage[all]{xy}
\usepackage{amsfonts}%
\usepackage{subfigure}

\definecolor{gray}{rgb}{0.4,0.4,0.4}

\def\({\left(}
\def\){\right)}
\def\[{\left[}
\def\]{\right]}
\def\<{\langle}
\def\>{\rangle}

\newcommand\half{{\ensuremath{\frac{1}{2}}}}
\newcommand\p{\ensuremath{\partial}}

\newcommand{\be}{\begin{equation}}
\newcommand{\ee}{\end{equation}}
\newcommand{\bea}{\begin{eqnarray}}
\newcommand{\eea}{\end{eqnarray}}
\newcommand{\bwt}{\begin{widetext}}
\newcommand{\ewt}{\end{widetext}}
\newcommand{\nn}{\nonumber\\}
\newcommand{\bi}{\begin{itemize}}
\newcommand{\ei}{\end{itemize}}
\newcommand{\ben}{\begin{enumerate}}
\newcommand{\een}{\end{enumerate}}
\newcommand{\bca}{\begin{cases}}
\newcommand{\eca}{\end{cases}}
\newcommand{\bln}{\begin{align}}
\newcommand{\eln}{\end{align}}
\newcommand{\bst}{\begin{split}}
\newcommand{\est}{\end{split}}

\newcommand\ep{\epsilon}

\newcommand\om{\omega}

\newcommand\ga{{\ensuremath{{\gamma}}}}

\newcommand\de{{\ensuremath{{\delta}}}}

\newcommand\ov{\over}
\newcommand\ha{{\half}}

\begin{document}

\title{Holographic Wilsonian RG flow  and Sliding Membrane Paradigm}
\author{Sang-Jin Sin$^{\dagger}$ and Yang Zhou$^{*}$}
\affiliation{$\dagger$ Department of Physics, Hanyang University, Seoul 133-791, Korea\\ $*$ Center for Quantum
Spacetime, Sogang University, Seoul 121-742, Korea}
\date{Feb.2011}

\vskip 2cm

\begin{abstract}

We study the relations between two different approaches to the holographic Renormalization Group (RG) flow
 at the dual gravity level:
One is the radial evolution of the classical equation of motion and the other is the flow equation given by the
holographic Wilsonian RG coming from the cut off independence.  Apparently, the two flows look different. We  give
general proofs that  the two flows are actually  equivalent. The role of the momentum continuity (MC)  is essential. We
show that MC together with cutoff independence gives the evolution equation of the boundary values. Equivalence of
conductivity flows in two paradigm has been shown as an explicit example.
 We also get the connecting  formula of Green functions
and  AC conductivity at arbitrary slice in terms of its value at horizon for various geometry backgrounds.
\end{abstract}
\maketitle

\section{Introdution}

From the early days of gauge/gravity dualities~\cite{Maldacena:1997re,Gubser:1998bc,Witten:1998qj},
   the role of radius as the energy scale was  emphasized explicitly and
   recent work on fermion dynamics in charged AdS$_4$~\cite{Faulkner:2009wj,Hartman:2009qu,Joe.Semiholography} also showed that
  the radial region trapping the fermions can play the role of fermi sea, demonstrating that the radial scale really encodes the
energy scale of the boundary dynamics. Therefore interpreting the  radial
 evolution of the classical solution as the renormalization group flow
  has been tempting and it was originally  encoded in the jargon ``scale radius duality''~\cite{kraus}  or
   ``UV/IR correspondence''~\cite{Susskind} and  it was  sharpened   in  ref. \cite{verlinde}, where
the flow was written in terms of   classical equation of motion and the Callan-Symanzik equation was written
in the context of AdS/CFT.

In the presence of the black hole, one can recast the problem in the following way.
There has been two different computations for  transport coefficients;   one at the horizon~\cite{son1} and the other
at the boundary~\cite{son2}.  To see the relations between the two holographic screens,
one may try to introduce a cutoff surface  at a finite radial position $r_c$
 and try to view  two holographic screens as  special cases of running cutoff surface.
In this framework, one can calculate the flow of physical quantities like conductivity \cite{Hong.Membrane} or
diffusion constant~\cite{Andy.RG} explicitly.  In this way,  the RG flow  can be quantitatively
formulated in finite temperature context, which is a nontrivial procedure in  field theory set up.

More recently, authors in the papers~\cite{Son.Deconstruction,Joe.RG,Hong.RG}, considered the Wilsonian renormalization
group flow~\cite{Wilson:RG} by dividing the radial direction into high (UV)  and low (IR) energy scales and integrating
out the UV part. As a result, the procedure results in  flow equations. The  question is whether the `traditional'
holographic RG flow coming from the classical equation of motion
 \cite{kraus,verlinde,Hong.Membrane,Andy.RG} is the same as the more recently formulated one in~\cite{Son.Deconstruction,Joe.RG,Hong.RG}.
 Partial answer to this question was already given in \cite{Joe.RG,Hong.RG} by showing that
solution to classical equation of motion can be used to construct the quadratic order of effective action.


In this paper, we point out that  holographic Wilsonian RG equation in gravity limit
is equivalent to flow equation coming from the equation of motion.
We first show this in general context and also  explicitly demonstrate it for the case of
 the conductivity. We will
also  derive an effective action with momentum correction up to second order.
 With a low frequency effective action at hand, we obtain the  relation of the  Green functions at two different
cut off surfaces, which can be viewed as the integrated version of the RG flow of the Green function.
Finally, we will show  that this relation precisely gives the double trace flow of Green
function if we integrate out UV geometry and it also precisely gives the transport coefficients flow which have been
obtained in sliding membrane paradigm if we integrate out IR geometry. We also give the solution of AC conductivity from a
constant value at horizon to another constant at the boundary. From the point of view in deconstructing holography, by
integrating out UV geometry, one obtained UV action of the effective coupling and a cutoff dependent dispersion
relation for Goldstone boson can be obtained.

\section{Sliding Membrane  and Holographic Wilsonian RG: a Review }

In this section, we  review some of the known results for RG flow  to set up our questions.
 We use the notation of ~\cite{Hong.Membrane,Son.Deconstruction}.
 The d+1
dimensional metric is $ds^2 = -g_{tt}dt^2 + g_{rr} dr^2 + g_{ij}dx^idx^j $.
We assume the above
metric has a horizon at $r=r_H$ and UV boundary at $r=\infty$. Except that $g_{tt}$ has a first order zero and $g_{rr}$
has a first order pole at the horizon, all other $g_{ij}$ are finite. $\{\mu,\nu\}$ run over $d$ dimensional space
time. We will use both $r$ and $z$ where  $r$ is used for the radial coordinate and $z$ is used for the inverse of it.
\vskip .2cm
 {\it Sliding Membrane: }
 Cutoff-dependent transport coefficients were defined on a sliding membrane in~\cite{Hong.Membrane}.
Such diffusion constant was
 worked out in
 ~\cite{Andy.RG}. In-falling horizon condition is shown to be equivalent
  to the regular horizon condition in Eddington-Finkelstein coordinates.
Let us take a bulk U(1) gauge field with  standard Maxwell action for example. All the quantities on the cutoff membrane can be
derived from an outer surface action  $S_{\rm b}=\int_{r_c}d^dx\ j^\mu A_\mu$, where $j^\mu \equiv \Pi^\mu = -\sqrt{-g}F^{r\mu}$.
Since one can visualize that
  the theory is defined on the cutoff surface which can be anywhere,  it can be called as a sliding membrane.
  That is, the source and the conjugated
momentum can be defined for any $r_c$ and $k_\mu$.
Assume the momentum is along the $i$ direction, then the longitudinal
$r_c$ dependent conductivity can be defined by $\sigma(r_c, k_\mu) = {j^i\ (r_c,k_\mu)\over F_{it}(r_c,k_\mu)}$.
From now on we delete subindex $c$ from $r_c$ if it is not confusing.
The equation of motion requests that $\sigma(r_c, k_\mu)$ satisfies the flow equation ~\cite{Hong.Membrane}:
\be\label{Mbraneflow} \small{\p_r\sigma = i\omega\sqrt{g_{rr}\ov
g_{tt}}\left[{\sigma^2\ov \Sigma(r)}\left(1-{k^2g^{i i}\ov \omega^2 g^{tt}}\right)-\Sigma(r)\right]\ ,\quad \Sigma(r) =
{\sqrt{-g}g^{ii}\ov \sqrt{g_{rr}g_{tt}}}\ .}\ee
The initial condition for this differential equation is not arbitrary since
regularity condition at the horizon requests that  $\sigma(r_H)= \Sigma(r_H)=
{\sqrt{-g} g^{ii} \over \sqrt{g_{rr} g_{tt} } }\biggr|_{r=r_H}$.
This result is  consistent with the definition  $\sigma(r_c, k_\mu) = {j^i\ (r_c,k_\mu)\over F_{it}(r_c,k_\mu)}$  applied at the horizon.
\vskip 0.2cm
 {\it Deconstructing Holography:}
In the work of \cite{Hong.Membrane,Andy.RG}, the dependence of physical quantities on the running cutoff implies the idea of  renormalization group in the context of the holography.
However, it was not very clear how the integrating out the high frequency degree of freedom  in the Wilsonian sense is implemented.
For this purpose, D.Nickel and D.T.Son~\cite{Son.Deconstruction} proposed a method to integrating out the UV dynamics.
The idea is the deconstruction:   replacing    the 5 dimensional UV action of $S=S_{\rm IR}(r<r_c)+S_{\rm UV}(r_c<r)$  by an effective 4 dimensional action: for the case of  Maxwell action,  the 4 dimensional action proposed was~\cite{Son.Deconstruction}  \be\label{Deconaction}  S_{\rm UV} = {1\ov 2}\int d^4x
\left[f_\mu\left(\p_\mu\varphi -a_\mu+ A_\mu \right)^2 \right] \ , \ee where
the goldstone field $\varphi =
\int_{r_c}^\infty A_r(r,x)$
is coming from the breaking the two $U(1)$ carried by the two boundary fields
$a_\mu=A_{\mu}(r_c)$ and $A_\mu=A_\mu(\infty)$ and
$f_\mu$ is given by metric factors. $a_\mu$is external source of the theory at $r_c$,  which is proposed to be determined from the current balance condition
\be \label{cbc} {\delta S\ov \delta a_\mu} = {\delta S_{\rm UV}\ov \delta a_\mu} +
{\delta S_{\rm IR}\ov \delta a_\mu}\equiv  j + j_{\rm IR} = 0\ . \ee
This equation together with the current conservation equation $\p_0j^0+\p_ij^i = 0$ and the definition of the conductivity  $\sigma\equiv { j_{\rm IR}\ov f_{i0}}={j\ov \p_0a_i-\p_ia_0}$
 gives the diffusion equation in the low
frequency limit $\p_0j^0 - {\sigma\ov f_t}\p_i\p^ij^0 = 0$.

At this moment, we want to pose a problem. The sliding membrane paradigm is to get the radial dependence of physical quantity from the
classical equation of motion and interpret the cut off dependence as the RG flow.
On the other hand, deconstruction is to obtain the radial dependence of physical quantity by integrating out the UV geometry. Are these two cut off dependence of transport coefficients consistent?
\vskip 0.2cm
 {\it Holographic Wilsonian RG}~~
In~\cite{Joe.RG}, it was suggested how to identify the bulk and boundary path integrals in Wilsonian approach where
 the whole boundary path integral is splited  into two parts: $$\small{Z=\int_{\rm boundary} DM_{k\delta<1}DM_{k\delta>1}e^{-S}=\int_{\rm boundary}
DM_{k\delta<1}e^{-S(\delta)}:=\left< \exp({-s(\delta))}\right>\ ,}$$
where $\delta$ is the cut-off length scale
and $S(\delta)=S_0
+ s(\delta)$.
One can derive RG equation from the  $\delta$ independence of the whole integral. The object of the Wilsonian RG
is to obtain $s(\delta)$.
The holographic version of this  is supposed to be
implemented by  splitting the radial scale into two and  write the bulk partition function accordingly:
$$Z=\int_{\rm bulk} D\phi_{z>\ep}D\phi_{z=\ep}D\phi_{z<\ep}e^{-S_{IR}(z>\ep)-S_{UV}(z<\ep)}=\int_{\rm bulk}
D\phi_{ \ep}\Psi_{IR}(\ep,{\tilde \phi})\Psi_{UV}(\ep,{\tilde \phi})$$
with ${\tilde \phi}=\phi(z=\ep)$. Notice that we are using inverse coordinate $z=1/r$ and $\ep=1/r_c$.
The question is how to identify the two path integral at the level of split?
The idea of \cite{Joe.RG} is that one should identify the $\Psi_{IR}(\ep,\phi_{z=\ep}) $ as the dual path integral with a UV cut off on a scale $\delta$ with a complete set of single trace operators and
the effective action of UV dynamics, $S_B=-\log \Psi_{UV}(\ep,\phi_{z=\ep})$,  can be determined  from the  independency of the cutoff or from the semiclassical approximation.
As a consequence, one can identify
\be
e^{-s(\delta)}=\int D{\tilde \phi} e^ { \int d^dx{\tilde \phi}(x){\cal O}(x) } e^{-S_B}.
\ee
 %

For later use, we review the U(1) gauge field example in $d+1$
dimensional fixed background  following \cite{Hong.RG}.
 The  action is given by $S = S_1 \left[z>\ep, A_M\right] + S_B[A_M,\ep]$, with $S_1 = -{1 \ov 4}
\int_{z> \ep}d^{d+1} x \sqrt{-g} \   F_{MN} F^{MN}$.
Here  $S_B$ is the boundary action containing   all
the information of integrating out UV physics.
The boundary momentum at $z=\ep$ for $S_1$ should satisfy the condition $\Pi_\mu\equiv-\sqrt{-g}F^{z\mu} = {\delta S_B\ov \delta A_\mu}$.
The presence of the boundary action $S_B$ is attributed to the multi-trace deformation ~\cite{Vecchi:2010dd, Joe.RG,
Hong.RG}. See~\cite{Witten:2001ua, Berkooz:2002ug, Mueck:2002gm, Minces:2002wp, Sever:2002fk,Li:2000ec} for earlier
discussions. The flow equation was worked out~\cite{Hong.RG} by requiring  $\frac{d }{d\ep} S=0$: \be
 \p_\ep S_B[{ A}_\mu,\ep] + H_1[A_\mu,{\de S_B \ov \de {A}_\mu }] =0\ ,
\label{HWRGflow} \ee
where $H_1$ is the hamiltonian of the IR action evaluated at $z=\ep$.
The  quadratic action (\ref{Deconaction})  is a solution satisfying flow equation of $S_B$ and
the corresponding effective action in boundary field theory
can be obtained by doing Legendre transform of $S_B$~\cite{Hong.RG}:  \be\label{beffaction} I_{\rm eff} = \int \left[
-{1\ov 2} \left({1\ov f_t}(j^0)^2-{1\ov f_i}(j^i)^2\right) -j^\mu(A_{b,\mu}+\p_\mu\varphi)\right]\ .
 \ee
We will see this boundary effective action can be  derived by  use of the deformed Green function later.

It is interesting to ask whether the presence of the $S_B$ can give any effect to the classical path at all.
In order to clarify these questions, we shall first generally demonstrate the equivalence between holographic Wilsonian RG flow in bulk classic level and radial flow of equation of motion. This also can solve our problem mentioned before for the relation between sliding membrane and deconstructing holography due to the same structure between deconstruction and holographic Wilsonian RG.


\section{Equivalence of holographic Wilsonian RG and Sliding Membrane paradigm}
There are two RG flows. One is the flow given by the classical equation of motion, and the other is the flow
from integrating out the geometry. Here we want to prove the equivalence of the two.
Start with the full quantum path integral in gravity side and do the  classical approximation
\be
Z_1[\phi_0]=\int_{\rm bulk} D\phi e^{-S}\simeq \exp(-S[\phi_{c}])
\ee
Here $\phi_0$ is the  value of bulk field  at the UV boundary  and
$\phi_H$ is a boundary condition at the horizon or at IR boundary. The latter can be
 a regularity condition at the horizon or infalling boundary condition for the probe field case.
 We emphasize that  it is  not necessarily a Dirichlet  boundary condition at the horizon.
$\phi_{c}$ is the solution of the classical equation of motion for  the given boundary condition.
Once two boundary conditions are specified classical field $\phi_c$ is defined for all $z$ and one can evaluate the
physical quantities at any slice. For example we can define the current and electric field and therefore the conductivity thereon. Such quantity will depends on the position of the slice and this is what we mean by the RG flow in sliding membrane.  Basically {\it the flow is given by the  bulk classical equation of motion}.

Now we introduce the splitting the path integral into UV and IR as before
\be
Z_2=\int_{\rm bulk} D{\tilde \phi}\Psi_{\rm IR}(\ep,{\tilde \phi})\Psi_{\rm UV}(\ep,{\tilde \phi}) \;,\ee
 and do classical approximation of each part:
\bea
\Psi_{\rm IR}(\ep,{\tilde \phi}) &=& \int_{\rm bulk} D\phi_{z>\ep}e^{-S(z>\ep)}=e^{-S_{ IR}[\phi_c^{IR}]}\\
\Psi_{\rm UV}(\ep,{\tilde \phi})   &=& \int_{\rm bulk} D\phi_{z<\ep}e^{-S (z<\ep)}=e^{-S_{ UV}[\phi_c^{UV}]},
 \eea
 where $S(z>\ep)=\int_\ep^{z_H}dz L[\phi]$ and similarly $S(z<\ep)=\int_0^\ep dz L[\phi]$.
Previously we wrote  $S_{\rm UV}$ as $S_B$ assuming that it can be expressed explicitly as  boundary values only.
It is easy to show that when the original action is quadratic in field variables,  $S_{UV}$ is actually quadratic form of the two boundary fields.
To appreciate the   meaning, we should be careful about the boundary condition (BC) of each field $\phi_c^{ IR}$ and
$\phi_c^{UV}$. The former is the   solution of  classical equation of motion of the action $S$ between $\ep<z<z_H$ with
the given  boundary conditions  $\phi_H$ and  ${\tilde \phi}$ at horizon and at $z=\epsilon$.
So it is better to write it as $\phi_c^{IR}[\phi_H,{\tilde \phi}]$.
The former is the solution with BC ${\tilde \phi},\phi_0$ at $z=\ep$ and $z=0$ respectively.
So we write it as $\phi_c^{UV}[{\tilde \phi},\phi_0]$.
Notice that so far ${\tilde \phi}$ is completely arbitrary and independent of  $\ep$.
Now we define
\be
S_\ep[\phi_H,{\tilde \phi}, \phi_0]=S_{IR}[\phi_c^{IR}[\phi_H,{\tilde \phi}]]+S_{UV}[\phi_c^{UV}[{\tilde \phi},\phi_0]].
\ee
Then we can write $Z_2$ as a functional integral  and take classical approximation again.
\be
Z_2=\int D{\tilde \phi}   e^{-S_\ep[{\tilde \phi}]}= e^{-S_\ep[ \phi_H,{\tilde \phi}^* , \phi_0]}
\ee
where  ${\tilde \phi}^*$ is the solution of the equation
\be\label{balance}
\frac{\delta S_\ep}{\delta{\tilde \phi}}= \frac{\delta S_{UV}}{\delta {\tilde \phi}} +\frac{\delta S_{IR}}{\delta {\tilde \phi}}=0, \ee
which is the condition of the minimizing the $S[{\tilde\phi}]$.
Since $\Pi=\pm\frac{\delta S}{\delta {\tilde \phi}}$ depending on whether  $\ep$ is upper or lower boundary of the
$z$ integral of the action, we have
\be
\Pi_{UV}=  \frac{\delta S_{UV}}{\delta {\tilde \phi}},\quad  \Pi_{IR}=- \frac{\delta S_{IR}}{\delta {\tilde \phi}}, \ee
which implies that the above eq. (\ref{balance}) is nothing but the continuity of momentum across the $z=\ep$.
\be
\Pi_{IR}=\Pi_{UV}.
\ee

What is the  solution  ${\tilde \phi}^*$?
Given the  boundary values $\phi_H$ and $\phi_0$, the bulk field $\phi(z)$ is determined for entire region  $z_H>z>0$
by the classical equation of motion, which will give the value of $Z_1$.
Notice that $\phi^*_{IR}$ (which is determined by $\phi_H$ and ${\tilde \phi}^*$)
should be joined with $\phi^*_{UV}$ (which is determined by   ${\tilde \phi}^*$ and $\phi_0$)
to give  a classical solution $\phi^*$.
  In order for $Z_1=Z_2$, it is sufficient to have
\be
{\tilde \phi}^*= \phi_c(z)\big |_{z=\ep} \hbox{ so that } \phi^*=\phi_c. \label{sol}
\ee
Notice that ${\tilde \phi}$  is originally defined as a boundary value at $z=\ep$ and it is independent of $\ep$.
We should also notice that
(\ref{sol}) guarantees the $\ep$ independence of the $Z_2$ and $S_\ep$, i.e,
$\frac{d S_\ep}{d \ep} =0 \label{epind},$
 because  with the solution  (\ref{sol}) $S_\ep$   becomes the $S[\phi_c]$, the classical value of action of $Z_1$
which is manifestly independent of $\ep$. So far we have shown that classical solutions satisfy the requirement of
$\ep$ independence of split partition $Z_2$. This is the reason why the solution of the Wilsonian RG equation can be
written  in terms of the solution of classical equation of motion.
\begin{figure}\centering
  \includegraphics*[bb=1 17 510 380,width=0.6\columnwidth]{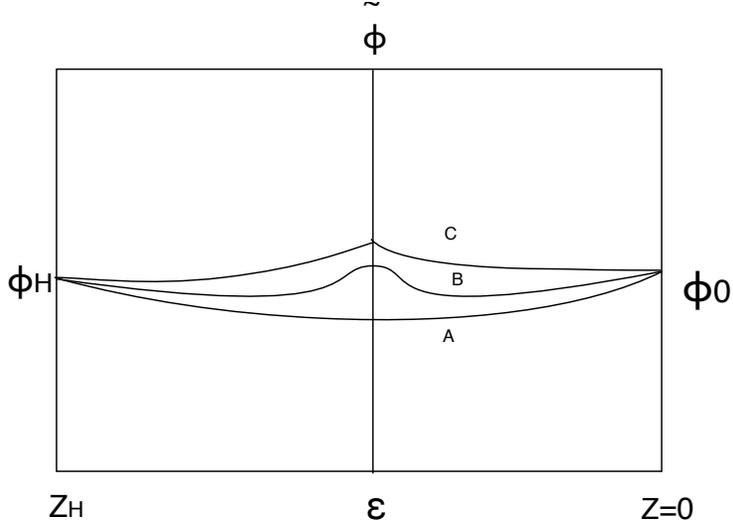}
  \caption{Flows with (A)  and without (C) the momentum continuity.
The uniqueness of the smooth solution forbids  a solution like B.  Therefore classical version of integration over
${\tilde \phi}$ pick up the ${\tilde \phi}=\phi_c(z)\big |_\ep$. The uniqueness of A means that classical flow of
$S_{UV}$ or $S_{IR}$ in $Z_2$ along $\ep$ can only pick up the path A.} \label{fig:equiv}
\end{figure}

Conversely, we want to argue that the continuity of the momentum (CM) implies that eq.(\ref{sol}) holds.
The CM means that the two solutions $\phi_{IR}$ and $\phi_{UV}$ are smoothly joined. Without CM, there is a solution
(with a cusp) which connects $\phi_H,\phi_0$ through $\tilde\phi$. See the curve C of Fig.1. Let's denote by  A   the
curve describing $\phi_c$ that smoothly connect the  given boundary conditions  $\phi_H$ and $\phi_0$, and suppose the
solution of
  the momentum continuity $\tilde \phi^*$ do not agree with  $\phi_c(\ep)$. Let's denote  by B the classical  path that
  connect   $\phi_H$ and $\phi_0$ through $\tilde \phi^*$.
The momentum (or velocity)  continuity   means that curve B in Fig.1  is also a smooth classical  path with the same boundary condition
of the curve A, which is contradiction to the uniqueness of the solution of differential equation.
The uniqueness of A means that classical flow of $S_{UV}$ or $S_{IR}$ in $Z_2$ along $\ep$ can only pick up the path A.
These arguments establish the equivalence of the classical radial flow and the Wilsonian RG flow.

\vskip .5cm
It is useful to see explicitly
how the `boundary conditions' $\tilde \phi$ at each slice  are patched together to agree
with the solution of radial equation of motion.
Actually it is the role of Wilsonian flow equation that
controls how the boundary condition $\tilde \phi$ should change
as one changes the membrane location.

So, we want to derive the radial evolution for $\tilde\phi$ from RG equation. For doing this we relax the $S_{IR}$ to
be off-shell by considering the action before doing path integral $D\phi_{IR}$ but after  $D\phi_{UV}$. Then, by
requesting the $\ep$ independence at this stage, one can easily establish the flow equation for the
$S_{UV}$~\cite{Joe.RG,Hong.RG} \be \label{HJE}
\partial_\ep S_{UV}[{\tilde \phi},\phi_0]+H[{\tilde \phi},  \Pi_{\tilde \phi} ] =0.
\ee $\Pi_{\tilde \phi}$ is  defined by $\Pi_{\tilde \phi}= \frac{\delta S_{UV}[{\tilde \phi},\phi_0]}{\delta {\tilde
\phi}}$ and so equal to $\Pi_{UV}$. The functional form of the Hamiltonian $H$ originally was a functional of the bulk
field $\phi (z)$ and its momentum conjugate $\Pi (z)$. It becomes functional of $\tilde \phi$ by evaluating it at
$z=\ep$ and it becomes a functional of $\Pi_{\tilde \phi}$ using the momentum continuity. Taking the derivative of eq.
(\ref{HJE}) with respect to ${\tilde \phi}$ we get \be \label{heq1}
 {\partial_\ep \Pi_{\tilde\phi } }= - \frac{\delta H}{\delta { \tilde\phi}}.
 \ee
The first term of this equation depends on   $\tilde \phi$   which is not $\ep$ dependent apparently so that the
equation can be a second order one.

The origin of the $\ep$ dependence of the `boundary value' $\tilde \phi$ is again coming from the
 momentum continuity.
The point is that while $\Pi_{UV}$ depends only on the boundary value $\tilde \phi$, $\Pi_{IR}$ is given by the
evaluation of the bulk field at $z=\ep$: $ \Pi_{IR}\equiv\frac{\delta L_{IR}}{\delta
{\dot\phi}}|_\ep=-\sqrt{-g}g^{zz}\p_z \phi(z)|_\ep$ for most of the second derivative Lagrangian
 so that the momentum continuity equation can also  be  written as
 \be \label{heq2}
 \Pi_{\tilde\phi}=-\sqrt{-g}g^{zz}\p_z \phi(z)\big |_\ep = -\sqrt{-g}g^{zz}\p_\ep\tilde\phi\ ,
  \ee
  giving the $\ep$ dependence of $\tilde \phi$.

Clearly (\ref{heq1}) and (\ref{heq2}) give the radial evolution for $\tilde\phi$. The equations (\ref{heq1}) and
(\ref{heq2}) together with $ \phi_0,\phi_H$ as the boundary condition of  ${\tilde\phi} $ repeat the whole classical
solution. This establishes the equivalence of the Wilsonian flow equation and the classical equation of motion. It
should be noticed that (\ref{HJE}) is usually used to find the form of effective action $S_B=S_{UV}$ but
 here its functional derivative is used to give the flow equation for the boundary value $\tilde \phi$.

The above proof goes through  for vector and tensor fields. Therefore we do not repeat the proof here.


Although   integrating out degree of freedom in terms of classical geometry,
these correspond to the quantum process in the boundary theory.
Integrating the geometry
gives deformation of the IR action in a  way determined by the  classical radial evolution,
which can be interpreted as the double trace deformation, whose dominance in turn  should be attributed to the large N nature.

\subsection{ Example: Flow of Conductivity }

In this subsection, we will point out the equivalence of holographic Wilsonian RG equation and the conductivity flow in
slicing membrane paradigm. We will derive the membrane flow equation of conductivity (\ref{Mbraneflow}) from the
Wilsonian RG equation (\ref{HWRGflow}). For simplicity, we start by assuming $S_B$ is quadratic form of $A_\mu$ and
consider only the longitudinal mode. In order to keep the same notation with~\cite{Hong.RG}, we use the metric \be ds^2
= -g_{tt}dt^2 + g_{ii}d\vec x^2 + g_{zz}dz^2\ .\ee $S_B$ was assumed as the form~\cite{Hong.RG} with $\hat{A}=A-\p\phi$  by
\begin{eqnarray}
 S_B [\hat A_\mu, \ep] = -  \ha \, \int\, {d^dk \over (2\pi)^d} \, \sqrt{-\ga}\; \left( f_0  (k,\ep) \hat A_0 (k) \hat A^0 (-k) +f_L (k, \ep)  g^{ii} \hat A^L (k)  \hat A^L (-k)
 \right)\ ,
 \label{HongSB}
\end{eqnarray} which can be rewritten in our language as\be\label{ourSB}
S_B=-  \ha \, \int\, {d^dk \over (2\pi)^d} \left(-G^{00}(k,\ep)\hat A_0 (k) \hat A_0 (-k) + G^{ii}(k,\ep) \hat A_i (k)
\hat A_i (-k)\right)\ , \ee where we assume the coefficient of mix term of $\hat A_0$ and $\hat A_i$ approximately vanishes ($f_{0i}\sim0$). From (\ref{HWRGflow}) with $A$ replaced by $\hat{A}$, by comparing coefficients one can obtain the flow equations for
$G^{00}$ and $G^{ii}$~\cite{Hong.RG}: \bea\label{Gflow1} \p_\ep G^{00} &=& - {(G^{00})^2\ov \sqrt{-g} g^{tt}g^{zz}} + \sqrt{-g} g^{tt}
g^{ii} k^2 \nn \p_\ep G^{ii} &=& - {(G^{ii})^2\ov \sqrt{-g} g^{ii}g^{zz}} - \sqrt{-g} g^{tt} g^{ii} \om^2\ . \eea The
definition of conductivity is given by \be \label{conductF} \sigma \equiv {J^i\ov E_i} = {J^i \ov -\p_0\hat A_i +
\p_i\hat A_0}\ . \ee With the definition of current $ J^\mu\equiv {\delta S_B\ov \delta A_\mu} $, from (\ref{ourSB}) we have \be -{1\ov
G^{00}} = {\hat A_0\ov J^0},\quad {1\ov G^{ii}} = {\hat A_i\ov J^i}\ . \ee Using conservation equation
$\p_0J^0+\p_iJ^i=0$, (\ref{conductF}) can be presented in momentum space as \be\label{sigmavsG} {1\ov \sigma} =
{i\ov \omega}\left({\omega^2\ov G^{ii}}-{k^2\ov G^{00}}\right)\ . \ee Using (\ref{Gflow1}) and (\ref{sigmavsG}), we
obtain \be\label{RGsigma} -{\p_\ep\sigma\ov i\omega} = \left(\sigma^2\left({1\ov \sqrt{-g}g^{ii}g^{zz}}-{k^2\ov
\omega^2}{1\ov \sqrt{-g}g^{tt}g^{zz}}\right)-\sqrt{-g}g^{tt}g^{ii}\right)\ , \ee where we used $g^{tt}g^{ii} k\omega\sim0$ which comes from the flow equation of approximately vanishing $f_{0i}$. Under coordinate transformation $z=1/r$, one has
 \be \p_\ep=-r^2\p_r,\quad  \sqrt{-g}_{(z)}g^{tt}g^{ii} = \sqrt{-g}_{(r)}g^{tt}g^{ii}\ r^2, \quad
  {1\ov \sqrt{-g}_{(z)}g^{zz}g^{ii}} = {1\ov \sqrt{-g}_{(r)}g^{rr}g^{ii}}\ r^2\ . \ee  we see that (\ref{RGsigma}) precisely gives the flow equation
(\ref{Mbraneflow}) in sliding membrane paradigm.

\subsection{RG Solutions: second order of momentum}

In this subsection, we want to find solutions for the RG equations (\ref{Gflow1}). We start by assuming
\be
{1\ov G^{00}} = {1\ov G^{00}_{(0)}} + k^2 {1\ov G^{00}_{(1)}}+ \cdots\ ,\quad {1\ov G^{ii}} = {1\ov G^{ii}_{(0)}} + \omega^2 {1\ov G^{ii}_{(1)}} + \cdots\
\ee
and
\be
\p_\ep G^{00}_{(0)} = - {(G^{00}_{(0)})^2\ov \sqrt{-g} g^{tt}g^{zz}}\ ,\quad\quad
\p_\ep G^{ii}_{(0)} = - {(G^{ii}_{(0)})^2\ov \sqrt{-g} g^{ii}g^{zz}}\ .
\ee
We have the following solutions by imposing simple vanishing boundary conditions for $G^{00}_{(1)}, G^{ii}_{(1)}$
\be
{1\ov G^{00}_{(1)}(z)} =\int_0^z -{ \sqrt{-g} g^{tt}
g^{ii} \ov G^{00}_{(0)}}\ ,\quad\quad
{1\ov G^{ii}_{(1)}(z)} = \int_0^z{\sqrt{-g} g^{tt}
g^{ii} \ov G^{ii}_{(0)}}\ .
\ee
We see that $k^2$ and $\omega^2$ correction are controlled by zero order solutions in the low frequency approximation.
\section{Flow Solutions From Integrating Out geometry}
Now we shall discuss the flow solutions for Green function coming from integrating out geometry over arbitrary region.

\subsection{Integrating out $z_0<z<\ep$ and   Effective $S_B$}
We
first derive an effective action by directly integrating out geometry of $z_0<z<\ep$ part, where $z=z_0$ and $z=\ep$
are two cutoff surfaces. We start from the Maxwell dynamics in the following $d+1$ dimensional general background
\be\label{zmetric} ds^2 = g_{tt}dt^2 + g_{ii}d\vec x^2 + g_{zz}dz^2\ , \ee with the metric components only depending on
$z$. Note that we can relate this metric to that in review section by $z={1\ov r}$ and we have no minus before $g_{tt}$
here. Maxwell equations are written by \be\label{EOM} \p_z\left[\sqrt{-g}\ (\p^z\,A^\mu - \p^\mu\,A^z)\right] +
\p_\nu\left[\sqrt{-g}\ (\p^\nu\,A^\mu - \p^\mu\,A^\nu)\right] = 0\ . \ee
We describe the
zero momentum limit case here and  leave the momentum corrected results in appendix A.
We assume
\be
A_\mu(z, k)=A^{(0)}_\mu(z)+ k^2A_\mu^{(1)}(z)+k_\mu k^\nu A_\nu^{(2)}(z) + \cdots
\ee and $A_z$ is independent on $k$. In the small momentum $\p\p\rightarrow
0$ limit, the first term will dominate and the above equation can be solved by
\be\label{zeroorder} \p_z \left[A^{(0)}_\mu - \p_\mu\int_{z_0}^z A_z dz\right] = {C_1^{\mu} \over
\sqrt{-g}g^{zz}g^{\mu\mu}}\ ,\ee
where $z=z_0$ is the initial position with $A_{\mu,z_0} = A^{(0)}_\mu(z_0)$ fixed. By
defining $ \varphi(x^\mu,z) = \int_{z_0}^z A_z dz$, the gauge invariant field
$\hat{A}^{(0)}_\mu = A^{(0)}_\mu - \p_\mu\varphi$
can  be solved in terms of the boundary condition of the gauge field at the $z=z_0$:
\be\label{IntegralS} \hat{A}^{(0)}_\mu(z) - A^{(0)}_\mu(z_0) = \int_{z_0}^z {C_1^{\mu} \over \sqrt{-g}g^{zz}g^{\mu\mu}} dz\
. \ee Remain task is to determine $C_1^{\mu}$.
At the other boundary $z=\ep$, we also need to give a  boundary condition. Since we know that we usually have to assign
a Dirichlet boundary condition at the infinite boundary as a rule of AdS/CFT correspondence (standard quantization), we
choose the same type of boundary condition at $\ep$ slice. Given  $\hat A^{(0)}_\mu (z=\ep)$,   $C_1^{\mu}$ is determined to
be
\be\label{solution2}  C_1^{\mu}= \frac{1}{\int_{z_0}^\ep {dz \over \sqrt{-g}g^{zz}g^{\mu\mu}}}    (\hat{A}^{(0)}_\mu (\ep) - A^{(0)}_{\mu ,z_0}  ):= {f_\mu}  (\hat{A}^{(0)}_\mu (\ep) - A^{(0)}_{\mu ,z_0}  )\,\ee
with
$ {1 \over f_\mu} =  \int_{z_0}^\ep {1 \over \sqrt{-g}g^{zz}g^{\mu\mu}} dz $  and $\mu$ runs over $d$ dimension.

\vskip 0.5cm

Now, we want to integrate out $z$ in region $z_0<z<\ep$. For  the standard quadratic Maxwell action, we obtain the
on-shell action as boundary term using the equations of motion:
 \be\label{SUV} S_{[z_0,\ep]}^{\rm on-shell} = - \half \int d^d x \sqrt{-g}g^{zz}g^{\mu\mu} \hat
A^c_\mu\p_z \hat A^c_\mu \biggr|^\ep_{z_0}\  = - \half  \int d^d x\ C_1^{\mu} \hat A^c_\mu\biggr|^\ep_{z_0}\ . \ee
where we used  $\p_z \hat A^{(0)}_\mu = {C_1^{\mu} \over \sqrt{-g}g^{zz}g^{\mu\mu}}$.
Using
solution (\ref{solution2}), we finally obtain the zero order on shell action \be\label{Effaction} S_{[z_0,\ep]}^{\rm on-shell} = - \half \int d^d
x\sum_\mu f_\mu( \hat{A}_\mu(\ep) - A_{\mu ,z_0} )(
\hat{A}_\mu(\ep) - A_{\mu,z_0} ).\ee

This expression was suggested first in~\cite{Son.Deconstruction} as UV effective action without proof. The same result
has been obtained for pure AdS case in~\cite{Joe.Semiholography}, while the above discussion works for general diagonal
backgrounds. The same result was given in~\cite{Hong.RG} without explicit derivation. We discuss in the general
diagonal geometry background with zero U(1) background charge, which also can contain the world volume of D brane.


\subsection{Green Function Flow}

With an effective action at hand, we want to find the relation between two Green functions defined at  the
two surfaces. In a background with a horizon at $z=z_H$, one  way to define the retarded Green functions at $z=\ep$
\cite{Hong.Membrane} is to assume that  the source and operator relation  is the same with  that at the $z=0$ which is
discussed in  \cite{Son.Realgreenfunction, Hong.Membrane}. The definitions of currents are given by
 \be J_{z_0}^\mu \equiv {\delta
S_{[z_0,\ep]}^{\rm on-shell} \ov \delta A_{\mu,z_0}} = J^\mu\ , \quad J_{\ep}^\mu \equiv {\delta S_{[z_0,\ep]}^{\rm
on-shell} \ov \delta A_{\mu,\ep}} = -J^\mu\ . \ee For simplicity, we only consider Green function in the diagonal  case in linear response \be
G_{z_0}^{\mu\mu} = {J_{z_0}^\mu\ov A_{\mu,z_0}}\ ,\quad G_{\ep}^{\mu\mu} = - {J_{\ep}^\mu\ov A_{\mu,\ep}}\ .\ee
 Since the current at left and right boundary of the $z_0<z<\ep$ zone is the same  in the
magnitude and opposite in direction, such $z$ independence of the current magnitude
gives a flow equation of the Green function.
The structure of the effective on shell action dictates that
Green function should satisfy \be\label{Gflow} {1\ov G_{\ep}^{\mu\mu}} - {1\ov G_{z_0}^{\mu\mu}} = {1\ov f_\mu} + {\p_\mu\varphi\ov J^\mu}\ .
\ee  It  describes the
holographic flow of Green function since both $z_0$ and $\ep$  can be  any places outside the horizon.

\subsubsection{$z_0=0$, $\ep <<z_H$, Related to Double Trace Flow}
When $z_0=0$ is fixed, the region we integrated becomes exactly the UV region and only the surface $z=\ep$ is left, which should
correspond to UV cut-off length scale in Wilson's description of quantum field theory by a certain way.
In the saddle point approximation, the whole bulk action becomes \be S = S_1 +
S_{[0,~\ep]}^{\rm on-shell}\ . \ee $S_{[0,~\ep]}^{\rm on-shell}$ can be treated as a boundary action on the bulk
dynamics of $z>\ep$ region. The quadratic boundary term is attributed to the double trace deformation term. The form of
Green function with double trace deformation can be generally written as~\cite{Hong.RG} \be\label{doubletrG} G_\kappa =
{1\ov G_{\kappa=0}^{-1} + \kappa}\ ,\ee where $\kappa$ is the double trace coupling and $G_{\kappa=0}$ is the Green
function without double trace deformation. In order to relate (\ref{doubletrG}) to our (\ref{Gflow}), note that when
$\kappa=0$, (\ref{doubletrG}) makes Green function well define at $z=0$ without double trace deformation. This is the
well-known Green function defined at $z=0$ in usual AdS/CFT, corresponding to the $z_0$ boundary Green function in our
discussion, which implies (up to renormalization) \be\label{G1} G_{\kappa=0} =
G_{z_0=0}^{\mu\mu}\ . \ee Along with it, the double trace deformed Green function can be given by (up to
renormalization) \be\label{G2} G_\kappa = G_\ep^{\mu\mu}\ . \ee Thus the effective action in boundary field theory can be restored from
\be\label{eff} \delta S_{\rm eff} =-{1\ov G_\kappa} \delta J\  J=-{1\ov G_\ep^{\mu\mu}} \delta J_\ep^\mu J_\ep^\mu\ . \ee where $J$ is field theory operator and $A$ is the corresponding source. Here we establish $J=J^\mu$ and $A=A_{\mu,\ep}$. Using (\ref{eff}) and (\ref{Gflow}) we obtain
\be
S_{\rm eff} = \int d^dx \sum_\mu\left[-\half(J^\mu)^2 {1\ov f_\mu}- J^\mu(A_{\mu,0}+\p_\mu\varphi)\right]\ .
\ee We see that it precisely gives the boundary
effective action (\ref{beffaction}) derived in~\cite{Hong.RG}, which provides a consistent check for the above discussions. Note that in order to keep the same direction with the current corresponding the usual boundary value of gauge field, it is natural to define $J^\mu=-J_\ep^\mu$ is the dual current operator in field
theory corresponding to source $A_{\mu,\ep}$. As a closing remark for this subsection, the flow equation given in
(\ref{Gflow}) can be related to the double trace deformation of the holographic Green function with help of (\ref{G1})
and (\ref{G2}).

\subsubsection{$z_0=0$, $\ep<z_H$, Related To Deconstruction}
In this subection we again fixed $z_0=0$, while keeping $\ep$ not far away from the
horizon. The reason is that we want to find long distance version of holographic liquid. We have proved in the low frequency limit, on shell action over $0<z<\ep$ is (\ref{Effaction}) with $z_0=0$,
which precisely gives the construction of $S_{\rm UV}$ in deconstructing holography~\cite{Son.Deconstruction}.

We start from the full action ($A_{\mu,b}=A_{\mu,0}$)\bea S &=& S_1 + S_{0<z<\ep}^{\rm  on-shell}\ . \eea Apparently, $S_1$ is the holographic part, which has a finite $\rm UV$
boundary $z=\ep$. To simplify the discussion, we assume $A_{\mu,b}$ vanishes at this moment. As usual real time
calculation in AdS/CFT, on shell $S_1$ equals to a boundary term at $\ep$ (with the gauge $A_z = 0$ and horizon
in-falling condition) \be + \half\int d^d x \sqrt{-g} g^{zz}A_\mu g^{\mu\mu}\p_zA_\mu\biggr|_\ep\ . \ee
With the help of $ G^{\mu\mu} = \sqrt{-g}~\p A^\mu/ A_\mu$ for the holographic part, we rewrite the action as \be
S_{\rm on-shell} = + {1\ov 4}\int_\ep d^d x ~G^{\mu\mu}A_\mu^2  - \half \int_\ep d^d x\sum_\mu f_\mu( A_\mu - \p_\mu
\varphi )^2\ . \ee
This action can be considered as the semi-holographic construction for
Maxwell fluctuations, an analogy with the semi-holographic Fermi liquid construction in~\cite{Joe.Semiholography}. From
this action, we can solve the dispersion relation for $\varphi$, the only excitation in the low frequency limit, once
we input correlation information in the holographic part. A novel example has been checked for holographic zero sound
in~\cite{Son.Deconstruction} in the limit $\ep\sim \infty$ at zero temperature. In principle, we can find the $\ep$
dependent dispersion relation, which may have more interesting applications.

To relate the Green function flow in this letter to the work~\cite{Son.Deconstruction} by D.Nickel and D.T.Son explicitly, note that the
conductivity defined on the $\ep$ slice in (\ref{defconduct}) can be equivalent to the IR conductivity $\sigma\equiv { j_{\rm IR}\ov f_{i0}}$ in~\cite{Son.Deconstruction}
 through a so called momentum balance condition. (Note that in our discussion, in order to get the flow
of Green function and conductivity, we do not need such a balance condition.) After dropping $E_{i,z_0}$, a diffusion
constant can be read from (\ref{flowconduct0})\be D = \sigma_\ep/f_0\ .\ee The only condition is
\be\label{Diffusioncondition} {\p_0J^i/ f_i} << {\p_iJ^0/ f_0}\ .\ee One observation in~\cite{Hong.RG} is that,
even we assume momentum always balance from the two sides of the cutoff surface $\ep$, the cutoff surface can not be
taken too closed to the horizon, since in that case the condition (\ref{Diffusioncondition}) is broken due to divergent
${1\ov f_i}$.
We shall study this in the following part.

\subsection{Conductivity Flow}

The flow of Green function depends on $\varphi$,  the massless Goldstone mode suggested  in~\cite{Son.Deconstruction}.
In the following, we will see the similar flow equation for conductivity  which depends only on the
background geometry. We start from the standard
definition of conductivity on two slices
\be\label{defconduct} \sigma_{z_0} = {J_{z_0}^i\ov E_{i,z_0}}\ ,\quad
\sigma_{\ep} = -{J_{\ep}^i\ov E_{i,\ep}} ,\ee
with $ E_i = - \p_0 A_i + \p_i A_0$. Using the the identity
$\p_0\p_i\varphi = \p_i\p_0\varphi$ we have \be\label{flowconduct0} {1\ov \sigma_{\ep}} - {1\ov \sigma_{z_0}} =
{- f_i^{-1}\p_0J^i + f_0^{-1}\p_iJ^0\ov J^i}\ . \ee With the conservation equation for $J^\mu$, one can easily rewrite the above
equation (\ref{flowconduct0}) in momentum space\be\label{flowconduct} {1\ov \sigma_{\ep}} - {1\ov \sigma_{z_0}} = - {ik^2 \ov \omega} {1\ov
f_0} - {i\omega \ov f_i}\ \quad \left(\p_i = ik,\ \p_0 =- i\omega\right).\ee In the diffusion region with small
$\omega\sim k^2$, (\ref{flowconduct}) becomes \be {1\ov \sigma_{\ep}} - {1\ov \sigma_{z_0}} = - {ik^2 \ov \omega} {1\ov
f_0}\ , \ee while in the region $k\rightarrow 0$ first, it becomes \be\label{Acflow} {1\ov \sigma_{\ep}} - {1\ov \sigma_{z_0}} =
{-i\omega \ov f_i}\ , \ee which is AC conductivity flow.  Apparently, (\ref{flowconduct}) gives the flow for
conductivity depending on small $k$ and $\omega$.

\subsubsection{$\ep\rightarrow z_H$, Related To Sliding Membrane Paradigm In Diffusion Region}

In this subsection, we will point out that flow of conductivity (\ref{flowconduct}) is equivalent to the sliding
membrane flow of conductivity in the diffusion region.
Within a diffusion scaling in the low frequency limit and the
initial condition at the horizon, the $r$ dependent conductivity can be solved by ~\cite{Hong.Membrane} \be\label{mconduct} {1\ov \sigma(r)}
= {1\ov \sigma_H }+ i{k^2\ov \omega} \int_{r_H}^r dr {1\ov \sqrt{-g}g_{rr}g_{tt}}\ , \ee

Apparently, we see that (\ref{flowconduct}) precisely give (\ref{mconduct}) in the low frequency diffusion region
$\omega\sim k^2$, where we should set \be z_0={1\ov r}\  , \quad \ep\rightarrow {1\ov r_H}\ , \ee and note that the integral in (\ref{mconduct}) is invariant under the coordinate transformation $z=1/r$.  In order to clarify
how we need to take the above limit, see (\ref{flowconduct}) again
and one can find that the only condition for correct\be {1\ov
\sigma_{z_0={1\ov r}}} - {1\ov \sigma_{\ep\rightarrow z_H}} = {ik^2 \ov \omega} {1\ov f_0} \ee is that \be {\omega \ov
f_i}<<1\ee holds in the limit $\ep\rightarrow z_H$, since in the diffusion region $\omega\sim k^2<<1$, ${k^2 \ov \omega}$ is
order one and ${1\ov f_t}$ is finite. It means that low $\omega$ should go to zero faster than the horizon limit. Using the horizon transport as an initial condition, one can obtain the running
diffusion constants on the sliding surface~\cite{Andy.RG}.

As a closing word at this moment, when we set $\ep\rightarrow {1\ov r_H}$ and $z_0$ finite, the $z_0$ slicing surface
is equivalent to the sliding membrane defined at $r$ in~\cite{Hong.Membrane}.

\subsubsection{Flow of AC conductivity}

We now turn to the flow of $\omega$ dependent conductivity with $k=0$.  For (\ref{Acflow}), we see that this formula is
not regular at the horizon because ${1\ov f_i}$ has divergence, thus it is not a complete RG formula for AC
conductivity. We will see that the complete RG formula should come from
(\ref{Mbraneflow}).
\begin{figure}[hbtp]
\includegraphics*[bb=0 236 594 610,width=0.45\columnwidth]{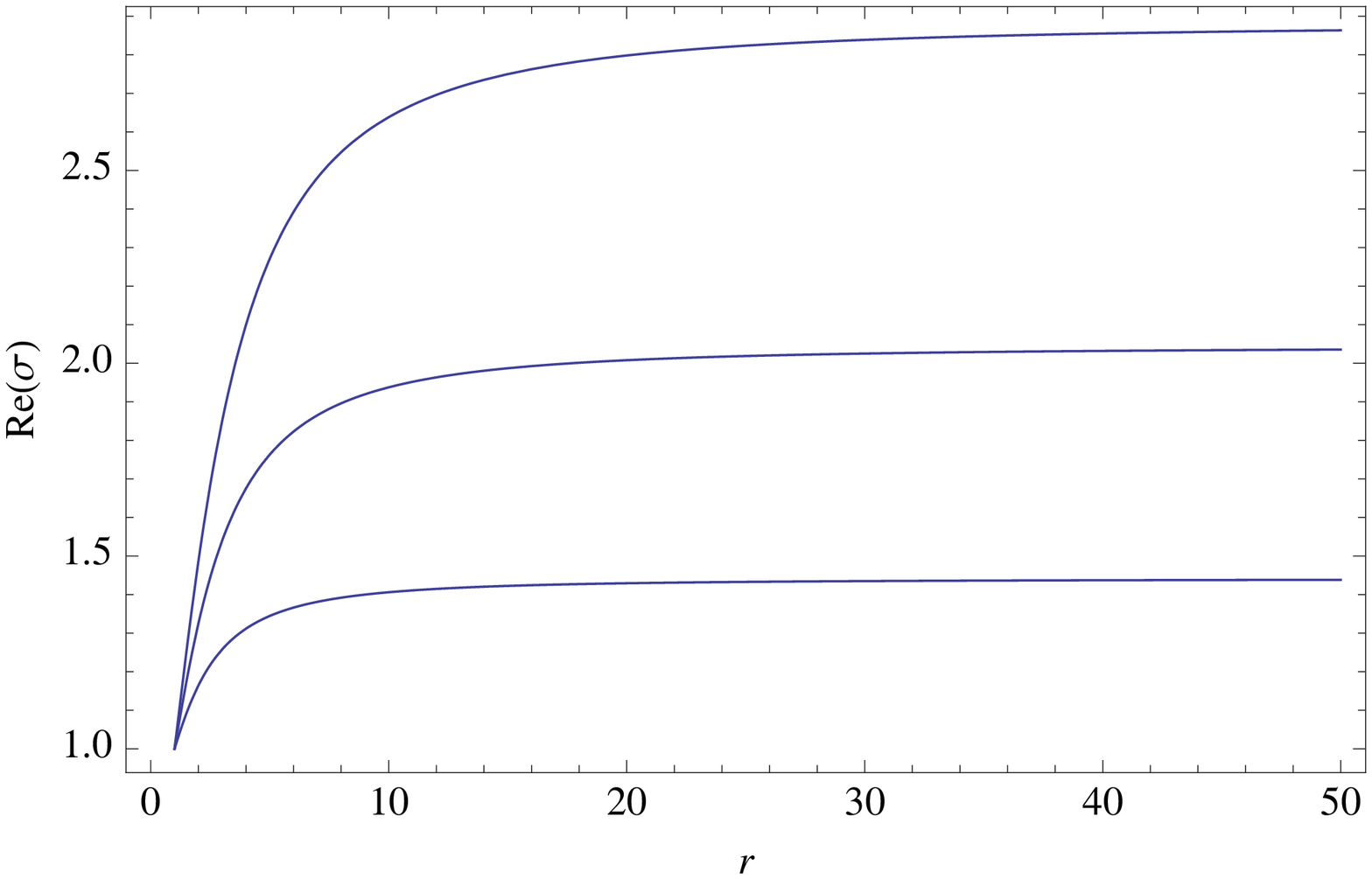}
\includegraphics*[bb=2 236 594 610,width=0.45\columnwidth]{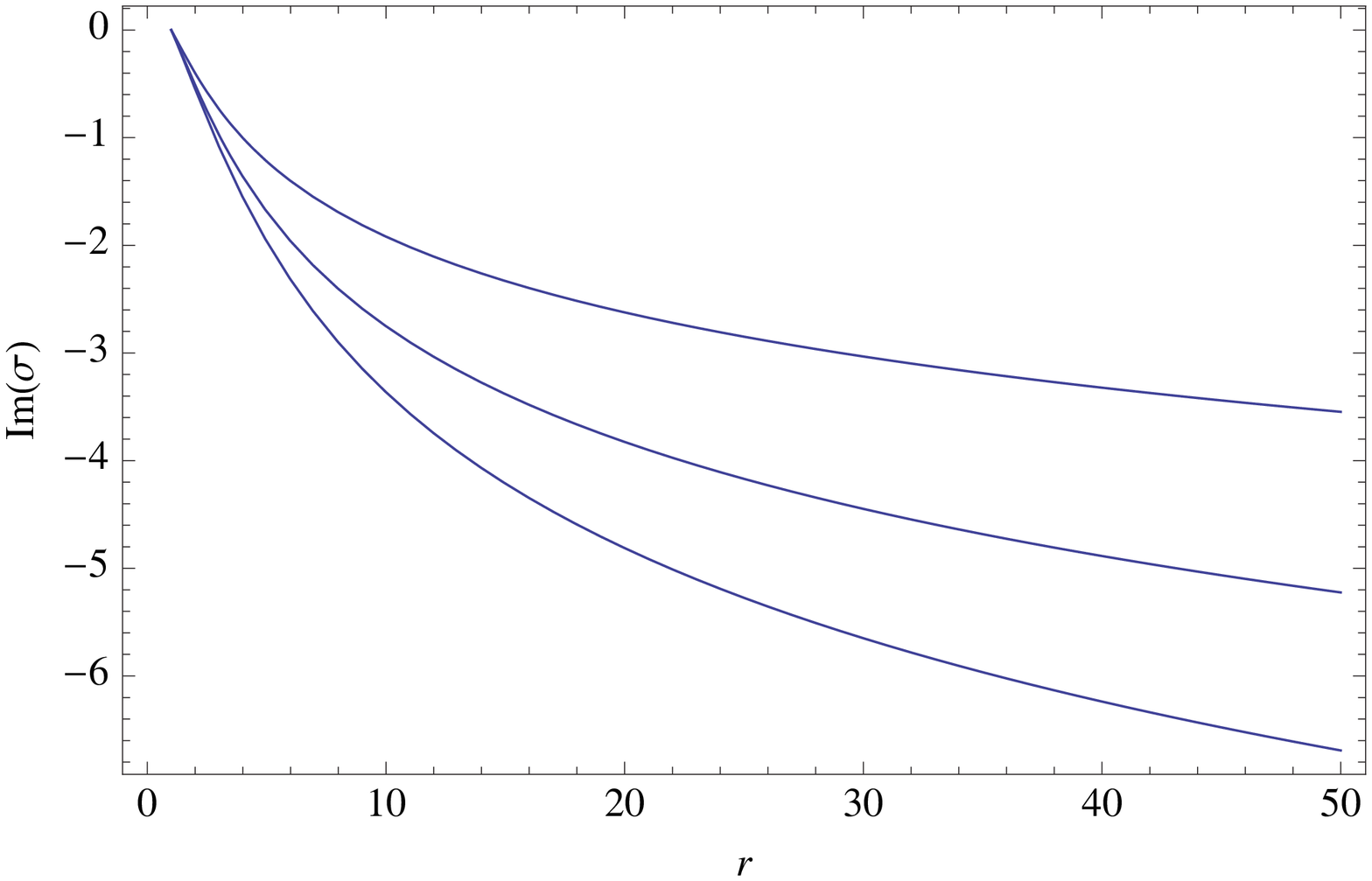}
\caption{ $r$ flow of AC conductivity, with $d=4$ AdS-black hole background and $\omega=2, 1.5, 1$, from up to down in the left Figure and inversely in
the right one. For $d>4$ AdS-black hole, the behavior of solutions are similar.
  \label{fig:RuningAcR}}
\end{figure}
After setting $k=0$, we have \be {\p_r\sigma\ov
i\omega} = {\sigma^2\ov \sqrt{-g}g^{rr}g^{ii}} - \sqrt{-g}g^{tt}g^{ii}\ . \ee
We
see that the above equation gives an exact flow for AC conductivity with the initial horizon value, which can be given by requesting equation regular at the horizon.
 We plot the $r$ flow function of AC conductivity for asymptotical AdS$_5$ black hole in Figure \ref{fig:RuningAcR}. For $d=3$, $r$ flow is trivial, and there is no $\omega$ dependence for boundary AC conductivity. For other cases, we show the result in the Figure \ref{fig2},\ref{fig3},\ref{fig4}.  For the Lifshitz black hole solutions for general $z$ and $d$, we refer to~\cite{AyonBeato:2010tm}.

 \begin{figure}[hbtp]
\includegraphics*[bb=2 242 603 592,width=0.45\columnwidth]{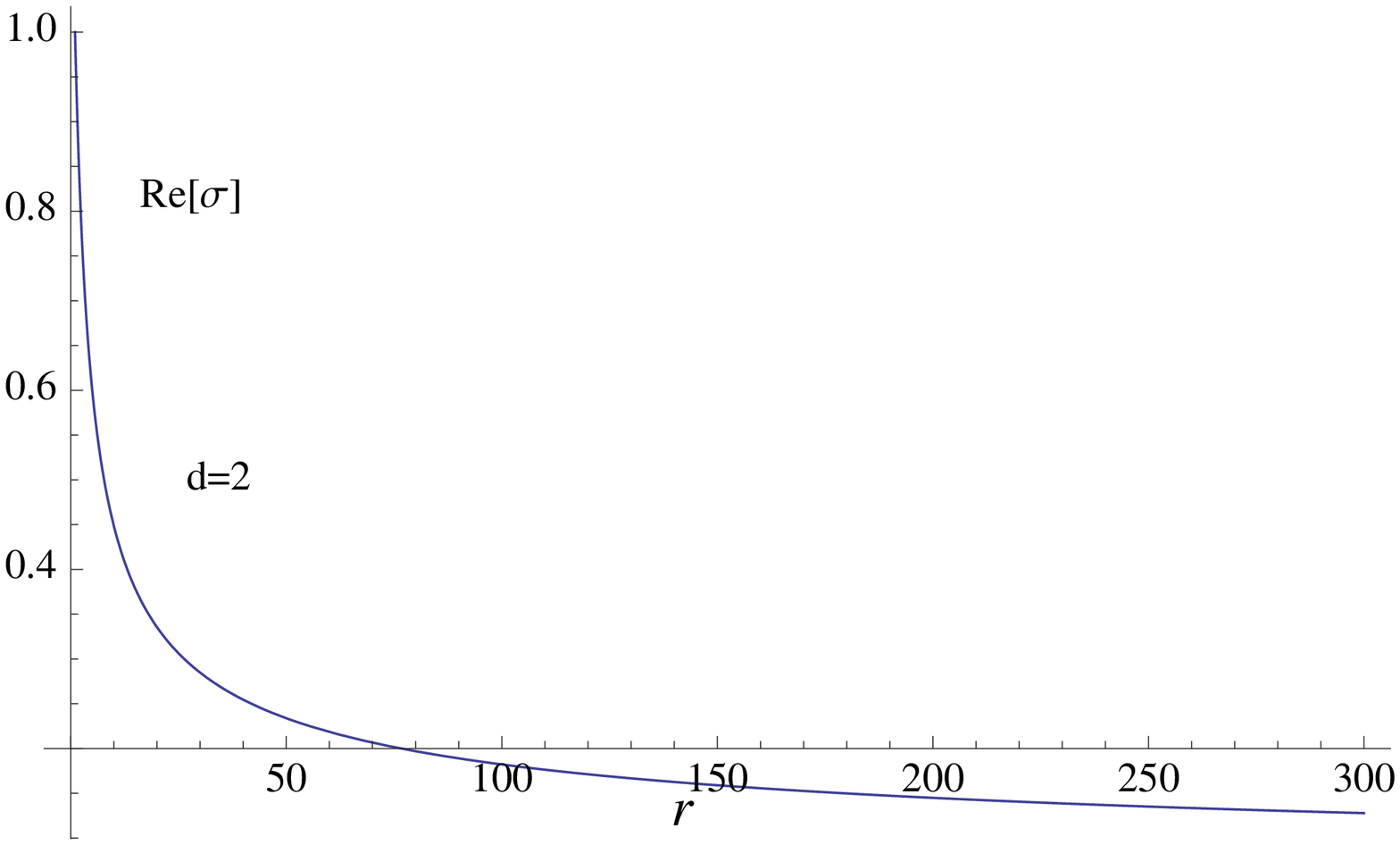}
\includegraphics*[bb=5 243 599 591,width=0.45\columnwidth]{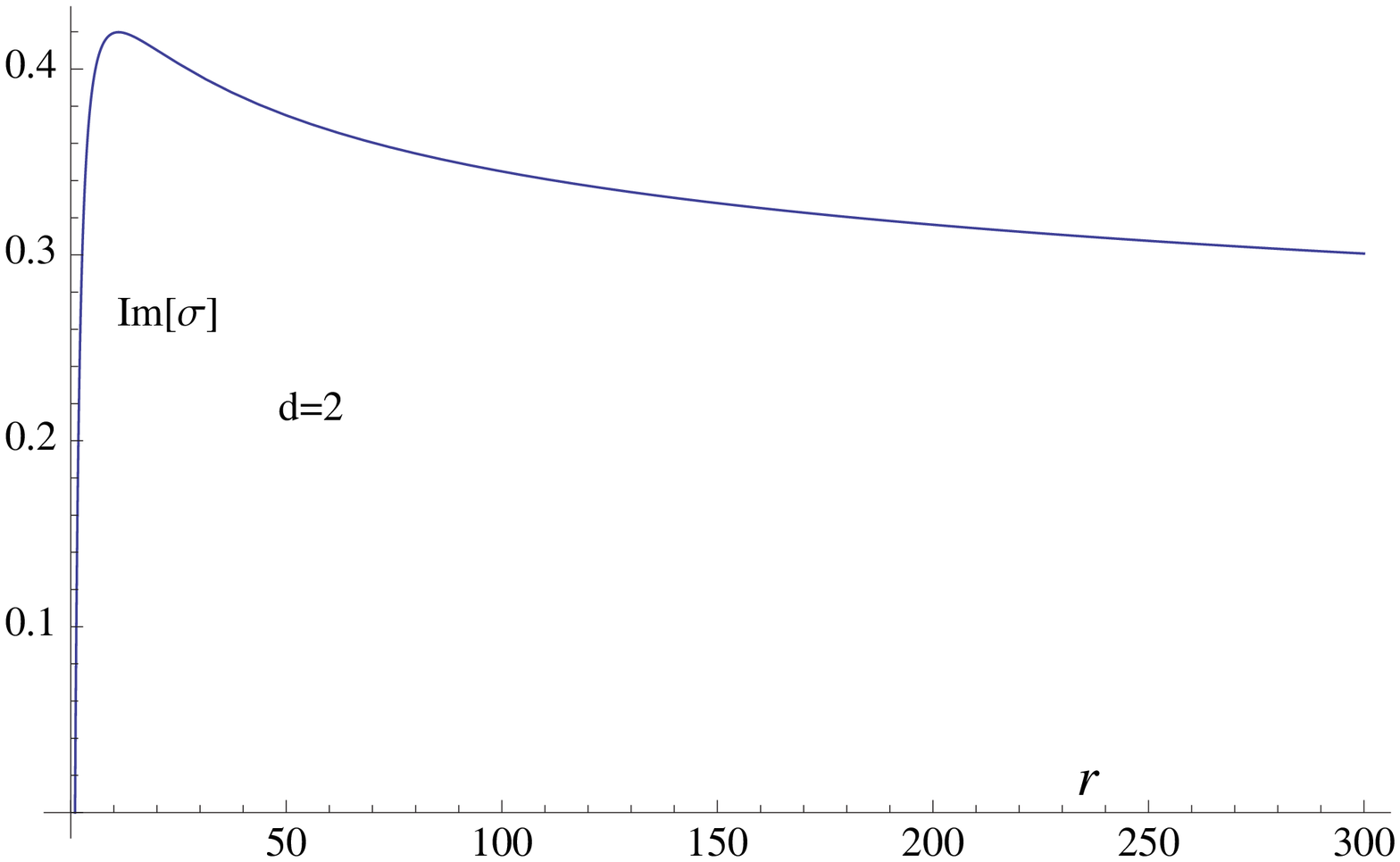}
\caption{ $r$ flow of AC conductivity, with $d=2$ AdS-black hole background, $d=1$ is similar.
  \label{fig2}}
\end{figure}

 \begin{figure}[hbtp]
\includegraphics*[bb=2 241 600 600,width=0.45\columnwidth]{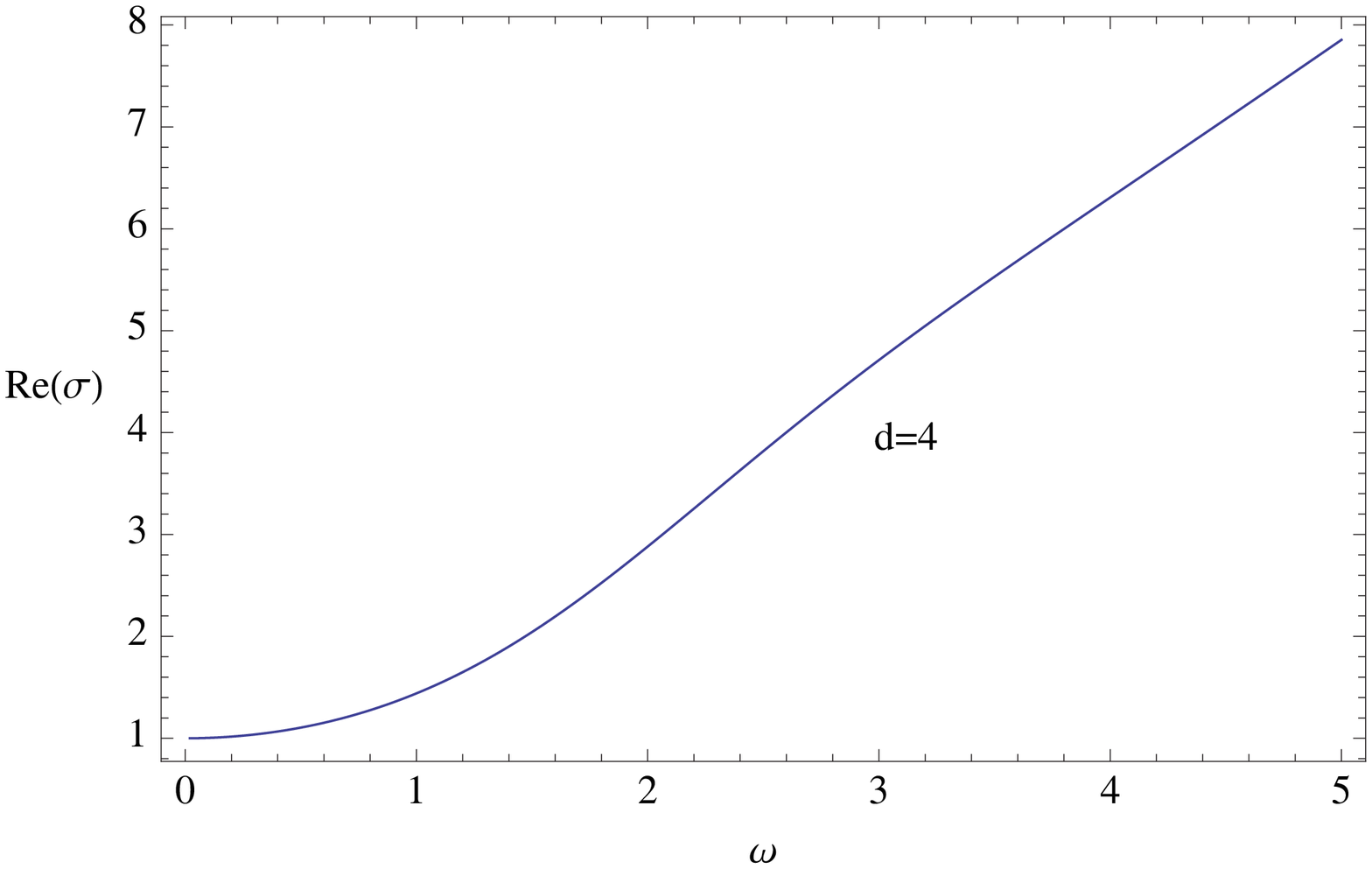}
\includegraphics*[bb=3 242 600 600,width=0.45\columnwidth]{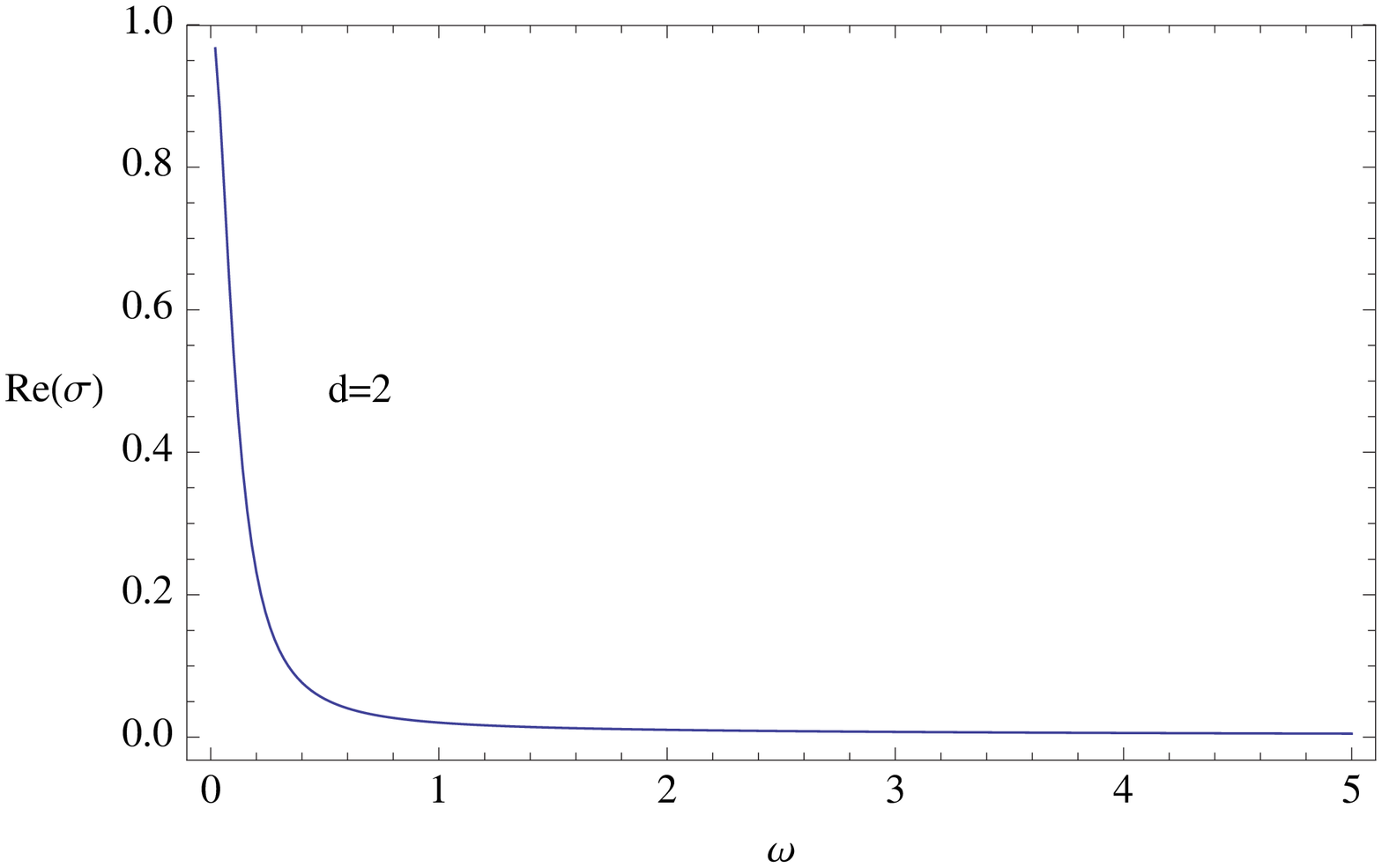}
\caption{ $\omega$ dependence of boundary AC conductivity, with $d=2,4$ AdS-black hole background. Behaviors of $d>4$ case are similar with $d=4$ and $d=1$ is similar to $d=2$.
  \label{fig3}}
\end{figure}

 \begin{figure}[hbtp]
\begin{center}
\includegraphics*[bb=6 245 600 610,width=0.45\columnwidth]{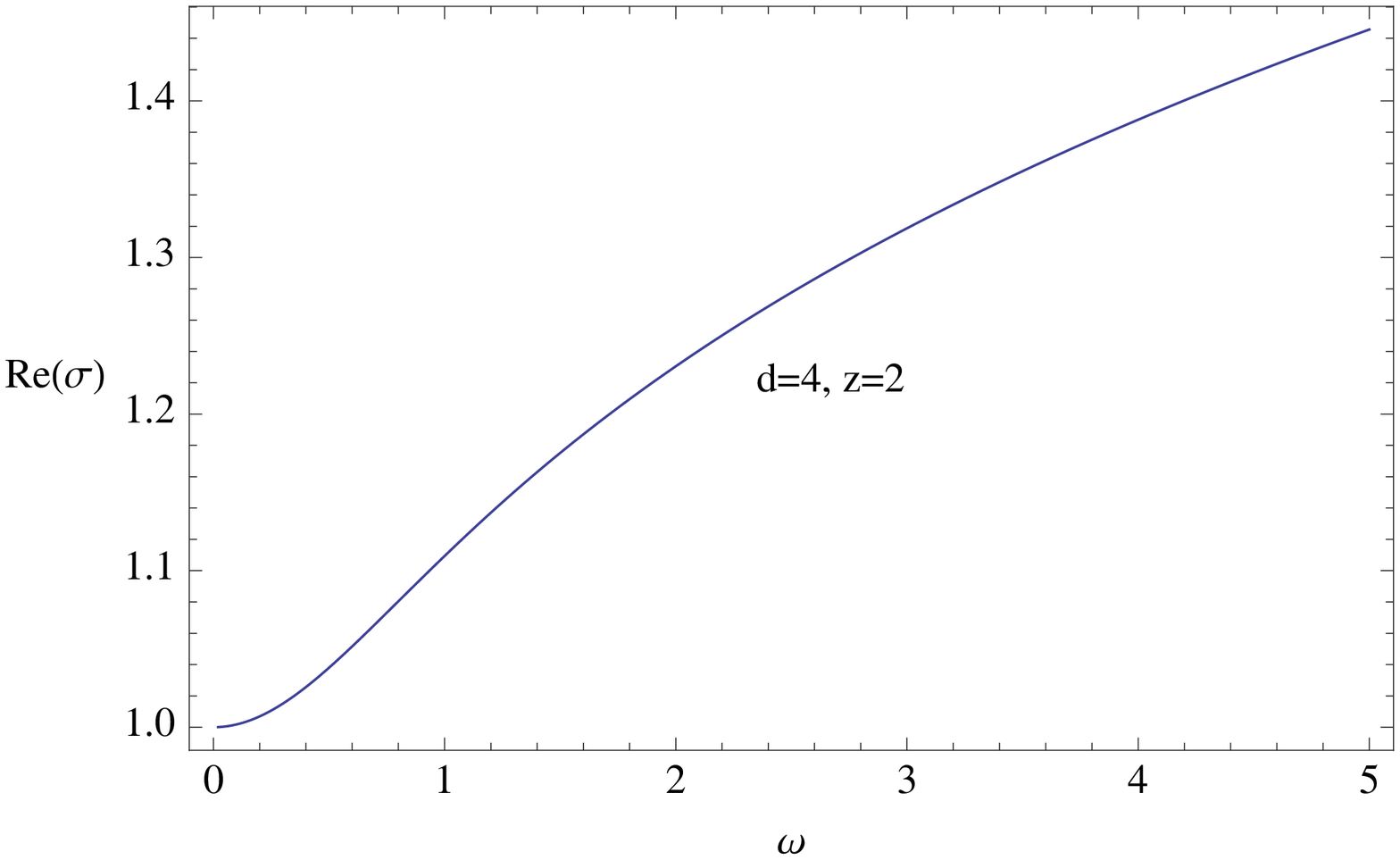}
\hspace{0.7cm}
\includegraphics*[bb=2 243 600 610,width=0.45\columnwidth]{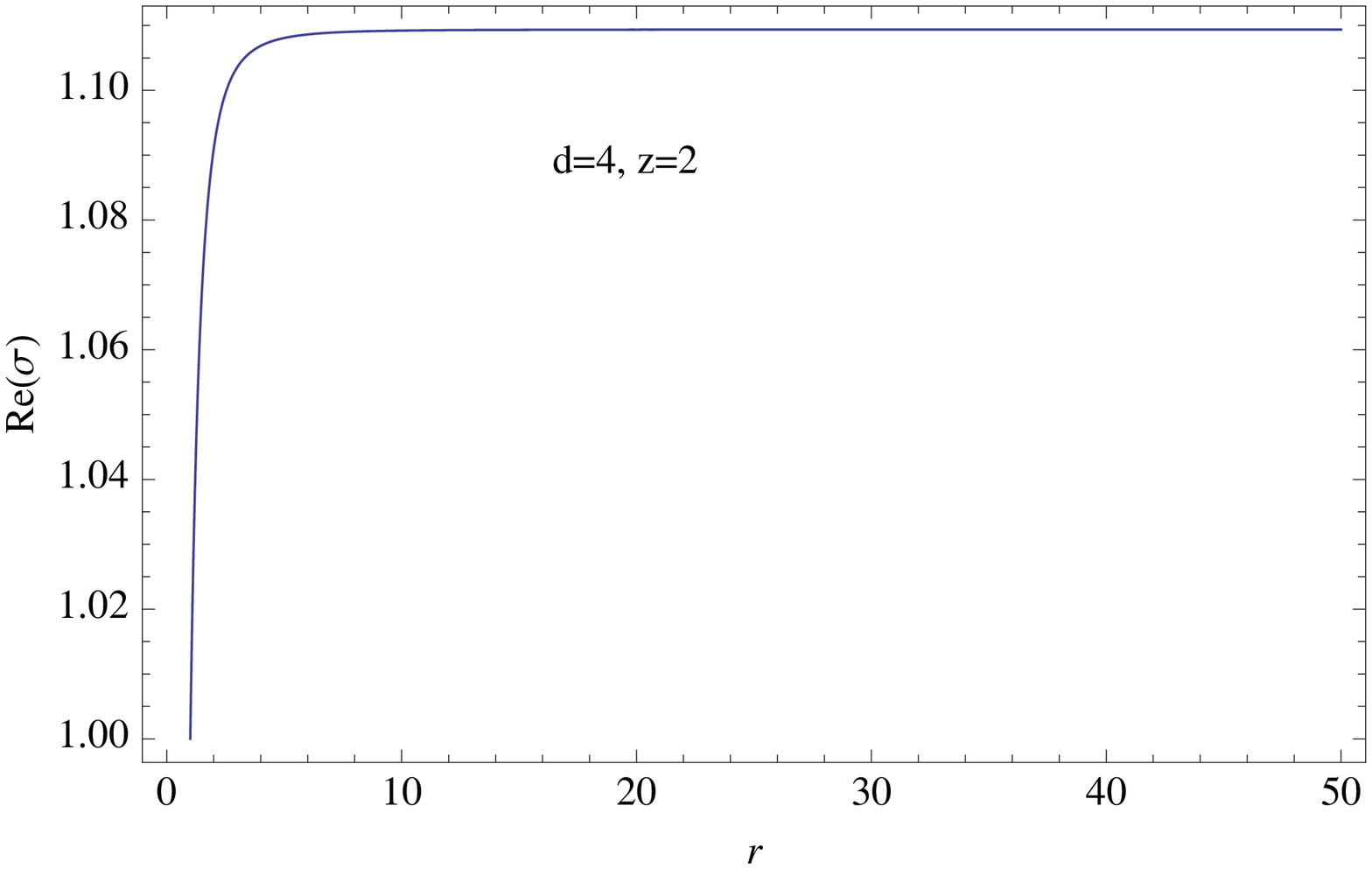}
\end{center}
\caption{ $\omega$ dependence of boundary AC conductivity and $r$ flow of AC conductivity, with $d=4$ and $z=2$ Lifshitz-black hole background~\cite{AyonBeato:2010tm}. Behaviors of $d>4$ case are similar with $d=4$.
  \label{fig4}}
\end{figure}

\section{Conclusion and Discussion}

To conclude, we would like clarify the relations among deconstructing holography, holographic Wilsonian RG flow and
sliding membrane paradigm as follows:
 Holographic Wilsonian RG equation in bulk classic level is equivalent to radial evolution of equation of motion.
The whole transport flow in
sliding membrane paradigm can be derived from holographic Wilsonian RG equation.
The lesson from the Wilsonian RG is that integrating the geometry
gives deformation of the IR action controlled by the classical bulk evolution, which can be interpreted as the
double trace deformation.

The method of integrating out geometry over arbitrary region $z_0<z<\ep$ can help to unify deconstructing holography,
holographic Wilsonian RG flow and sliding membrane paradigm. $S_{\rm UV}$ part in deconstructing holography can be
derived by integrating out UV geometry in low frequency limit. The running dispersion relation of Goldstone boson can
be obtained once we know IR holographic correlation precisely. The effective interacting action with finite momentum
corrections up to second order has been obtained. In the zero frequency limit, for the UV part $0<z<\ep$, current
$J^\mu$ is a constant and the flow of deformed Green function can be derived from the quadratic effective action, which
is consistent with the double trace formula of holographic Green function. When we set $\ep\rightarrow {1\ov r_H}$ and
$z_0$ finite, the $z_0$ sliding surface is equivalent to the sliding membrane. In the diffusion region,  conductivity
flow from integrating out geometry method is equivalent to that in sliding membrane paradigm.


\section*{Acknowledgements}
YZ would like to thank Prof. Hong Liu for simulating discussions in the KITPC workshop. SJS thanks  Shesansu Pal and Song He
for useful discussions. This work was supported by the National Research Foundation of Korea(NRF) grant funded by the
Korea government(MEST) through the Center for Quantum Spacetime(CQUeST) of Sogang University with grant number
2005-0049409. SJS was also supported by Mid-career Researcher Program through NRF grant (No. 2010-0008456 ).

\appendix
\section{Effective action with finite momentum correction}
In order to consider the finite momentum correction, we need to solve (\ref{EOM}) without dropping the term with second order derivative. Since the full Maxwell equation is hard to solve in the curved space time we consider the second order derivative only as correction. For convenience, we work in momentum space and  take the momentum to be along the $i$ direction. Then Maxwell equations separate into two groups: longitudinal channel and transverse channel. Let us first consider the longitudinal only, then the equation of motion (\ref{EOM}) involves
\bea
\p_z[\sqrt{-g}g^{zz}g^{00}F_{z0}] + \sqrt{-g}g^{ii}g^{00}\p_iF_{i0} = 0\ ,\\
\p_z[\sqrt{-g}g^{zz}g^{ii}F_{zi}] + \sqrt{-g}g^{ii}g^{00}\p_0F_{0i} = 0\ .
\eea
We expand the $A_0$ and $A_i$ on small momentum as
\bea\label{Aexpansion}
A_0 = A^{(0)}_0(z) + k_i^2A^{(1)}_0(z) + k_i\omega A_0^{(2)}(z) +\cdots \nn
A_i = A^{(0)}_i(z) + \omega^2A^{(1)}_i(z) + k_i\omega A_i^{(2)}(z) +\cdots
\eea
 { The term linear in $k$ and $\omega$ is deleted in the above expansion due to the structure of equations of motions.}
Substituting (\ref{Aexpansion}) into equations of motion and using the zero-th order solution (\ref{zeroorder}), one can obtain the following equations for $A_0^{(1)}$, $A_0^{(2)}$, $A_i^{(1)}$, $A_i^{(2)}$ by dropping higher momentum corrections (more than cubic) and assuming $k$ and $\omega$
are independent parameters\bea\label{firstorder}
\p_z[\sqrt{-g}g^{zz}g^{00}\p_z A_0^{(1)}] - \sqrt{-g}g^{ii}g^{00}A_0^{(0)} = 0\\
\p_z[\sqrt{-g}g^{zz}g^{00}\p_z A_0^{(2)}] - \sqrt{-g}g^{ii}g^{00}A_i^{(0)} = 0\nn
\p_z[\sqrt{-g}g^{zz}g^{ii}\p_z A_i^{(1)}] - \sqrt{-g}g^{ii}g^{00}A_i^{(0)} = 0\nn
\p_z[\sqrt{-g}g^{zz}g^{ii}\p_z A_i^{(2)}] - \sqrt{-g}g^{ii}g^{00}A_0^{(0)} = 0 .
\eea
Notice that we have worked out the solutions for $A_0^{(0)}$ and $A_i^{(0)}$, one can easily write the solutions for $A_0^{(1)}$, $A_0^{(2)}$, $A_i^{(1)}$, $A_i^{(2)}$ by solving the above equations.
From (\ref{IntegralS}) we have
\be
A_0^{(0)}(z) = A_0^{(0)}(z_0) + \int_{z_0}^z {C_1^0\over \sqrt{-g}g^{zz}g^{00}} dz +\p_0\varphi\ ,
\ee
By defining the function
\be
F_0(z) = \sqrt{-g}g^{ii}g^{00} \left[A_0^{(0)}(z_0) + \int_{z_0}^z {C_1^0\over \sqrt{-g}g^{zz}g^{00}} dz\right]\ ,
\ee
one can solve (\ref{firstorder}) by
\be
\sqrt{-g}g^{zz}g^{00}(\p_zA_0^{(1)}(z)) = \int_{z_0}^z dzF_0(z) + C_2^{00}\ ,
\ee
where $C_2^{00}$ can be determined properly. The final solution for $A_0^{(1)}$ can be given by
\be\label{Integral2}
A_0^{(1)}(z) = \int_{z_0}^z {\int_{z_0}^z F_0(z)dz + C_2^{00}\over \sqrt{-g}g^{zz}g^{00}} + A_0^{(1)}(z_0)\ .
\ee
$A_0^{(2)}$, $A_i^{(1)}$, $A_i^{(2)}$ can be determined by the same way
\bea
A_0^{(2)}(z) = \int_{z_0}^z {\int_{z_0}^z F_i(z)dz + C_2^{i0}\over \sqrt{-g}g^{zz}g^{00}} + A_0^{(2)}(z_0)\\
A_i^{(1)}(z) = \int_{z_0}^z {\int_{z_0}^z F_i(z)dz + C_2^{ii}\over \sqrt{-g}g^{zz}g^{ii}} + A_i^{(1)}(z_0)\\
A_i^{(2)}(z) = \int_{z_0}^z {\int_{z_0}^z F_0(z)dz + C_2^{0i}\over \sqrt{-g}g^{zz}g^{ii}} + A_i^{(2)}(z_0)\ ,
\eea
where the definitions for $F_i(z)$ is given by
\be
F_i(z) = \sqrt{-g}g^{ii}g^{00} \left[A_i^{(0)}(z_0) + \int_{z_0}^z {C_1^i\over \sqrt{-g}g^{zz}g^{ii}} dz\right] +\p_i\varphi\ .
\ee
Just as we determine $C_1^\mu$ in (\ref{solution2}), we want to determine the $C_2^{00}$, $C_2^{i0}$, $C_2^{ii}$, $C_2^{0i}$ by the boundary values of gauge field and  factors involving integrating out geometry. We use (\ref{Integral2}) to determine $C_2^{00}$, which is obtained as
\be
C_2^{00} = \frac{1}{\int_{z_0}^\ep {dz \over \sqrt{-g}g^{zz}g^{00}}}   \left({A}^{(1)}_0 (\ep) - A^{(1)}_{0 ,z_0} - \int_{z_0}^\ep {\int_{z_0}^z F_0(z)dz \over \sqrt{-g}g^{zz}g^{00}}\right)\ .
\ee
In the same way, $C_2^{i0}$, $C_2^{ii}$, $C_2^{0i}$ can be obtained as follows
\bea
C_2^{i0} &=& \frac{1}{\int_{z_0}^\ep {dz \over \sqrt{-g}g^{zz}g^{00}}}   \left({A}^{(2)}_0 (\ep) - A^{(2)}_{0 ,z_0} - \int_{z_0}^\ep {\int_{z_0}^z F_i(z)dz \over \sqrt{-g}g^{zz}g^{00}}\right)\\
C_2^{ii} &=& \frac{1}{\int_{z_0}^\ep {dz \over \sqrt{-g}g^{zz}g^{ii}}}   \left({A}^{(1)}_i (\ep) - A^{(1)}_{i ,z_0} - \int_{z_0}^\ep {\int_{z_0}^z F_i(z)dz \over \sqrt{-g}g^{zz}g^{ii}}\right)\\
C_2^{0i} &=& \frac{1}{\int_{z_0}^\ep {dz \over \sqrt{-g}g^{zz}g^{ii}}}   \left({A}^{(2)}_i (\ep) - A^{(2)}_{i ,z_0} - \int_{z_0}^\ep {\int_{z_0}^z F_0(z)dz \over \sqrt{-g}g^{zz}g^{ii}}\right)\ .
\eea
Now we shall write down the full effective action containing the finite momentum corrections. Note that  the first equal sign in (\ref{SUV}) holds even at high momentum corrections, one can substitute the solutions $A_0^{(1)}$, $A_0^{(2)}$, $A_i^{(1)}$, $A_i^{(2)}$ into (\ref{Aexpansion}) and evaluating (\ref{SUV}). We write the effective action by dropping the higher order momentum corrections as follow
\bea
S^{\rm onshell} = S_{[z_0,\ep]}^{\rm on-shell} + C_1^0\left[k_i^2(A_0^{(1)}(\ep)-A_{0,z_0}^{(1)})+k_i\omega(A_0^{(2)}(\ep)-A_{0,z_0}^{(2)})\right]\nn
+ k_i^2\left[\int_{z_0}^\ep F_0(z)dz\times \hat A_0^{(0)}(\ep) + C_2^{00}(\hat{A}^{(0)}_0 (\ep) - A^{(0)}_{0 ,z_0} ) \right]\nn
+ k_i\omega\left[\int_{z_0}^\ep F_i(z)dz\times \hat A_0^{(0)}(\ep) + C_2^{i0}(\hat{A}^{(0)}_0 (\ep) - A^{(0)}_{0 ,z_0} ) \right]\nn
+C_1^i\left[\omega^2(A_i^{(1)}(\ep)-A_{i,z_0}^{(1)})+k_i\omega(A_i^{(2)}(\ep)-A_{i,z_0}^{(2)})\right]\nn
+\omega^2\left[\int_{z_0}^\ep F_i(z)dz\times \hat A_i^{(0)}(\ep) + C_2^{ii}(\hat{A}^{(0)}_i (\ep) - A^{(0)}_{i ,z_0} ) \right]\nn
+ k_i\omega\left[\int_{z_0}^\ep F_0(z)dz\times \hat A_i^{(0)}(\ep) + C_2^{0i}(\hat{A}^{(0)}_i (\ep) - A^{(0)}_{i ,z_0} ) \right]\ .
\eea
This action describes the effective coupling between $A_\mu(\ep)$, $A_{\mu, z_0}$ and $\varphi$  including momentum correction up to second order.
Flow solutions for transport coefficients from this effective action should be consistent with solution solved from the classical equations of motion under low frequency approximation.

\end{document}